\documentclass[journal]{IEEEtran}

\usepackage[utf8]{inputenc}
\usepackage{amsmath}
\usepackage{cite}
\usepackage{multicol}
\usepackage{graphicx}
\usepackage{array}
\usepackage{multirow}
\usepackage{graphicx}
\usepackage{enumitem}
\usepackage{color}
\usepackage{nomencl}
\usepackage{algorithm}
\usepackage{algpseudocode}
\usepackage{ifthen}
\usepackage{etoolbox}
\usepackage{amssymb}
\usepackage{comment}
\usepackage{mathrsfs}
\usepackage{mathtools}
\usepackage{bbold}
\usepackage{yfonts}
\usepackage{soul}
\usepackage{placeins}
\usepackage{fnpct}
\usepackage{courier}
\usepackage{ntheorem}
\usepackage{cuted}
\usepackage{stfloats}
\usepackage{hyperref}
\usepackage{academicons}

\theoremseparator{:}

\newtheorem{hyp}{Hypothesis}

\usepackage{fnpct}
\ifCLASSOPTIONcompsoc \usepackage[caption=false,font=normalsize,labelfon
t=sf,textfont=sf]{subfig}
\else
\usepackage[caption=false,font=footnotesize]{subfig}
\fi
\usepackage{tikz}
\usetikzlibrary{decorations.pathreplacing}
\usetikzlibrary{arrows,shapes,positioning}
\usetikzlibrary{patterns}
\usepackage{pgfplots}
\pgfplotsset{compat=newest}
\pgfplotsset{plot coordinates/math parser=false}
\newlength\figureheight
\newlength\matlabfigurewidth
\newcommand\moverbar[1]{\bar{#1}}

\newcommand\Item[1][]{%
  \ifx\relax#1\relax  \item \else \item[#1] \fi
  \abovedisplayskip=0pt\abovedisplayshortskip=0pt~\vspace*{-\baselineskip}}

\title{\LARGE Quantifying the Uncertainty of Sensitivity Coefficients Computed from Uncertain Compound Admittance Matrix and Noisy Grid Measurements}
\author{
    \IEEEauthorblockN{
    Rahul Gupta\thanks{Rahul Gupta is with Distributed electrical systems laboratory, École polytechnique fédérale de Lausanne (EPFL),
1015 Lausanne, Switzerland (e-mail: rahul.gupta@epfl.ch) Orcid: \href{https://orcid.org/0000-0002-9905-6092}{0000-0002-9905-6092}.}, \emph{Member IEEE}\\}
   }

\usepackage{etoolbox}
\makeatletter
\patchcmd{\@maketitle}
  {\addvspace{0.5\baselineskip}\egroup}
  {\addvspace{-1.5\baselineskip}\egroup}
  {}
  {}
\makeatother
\begin{document}
\maketitle
\begin{abstract}
The power-flow sensitivity coefficients (PFSCs) are widely used in the power system for expressing linearized dependencies between the controlled (i.e., the nodal voltages, lines currents) and control variables (e.g., active and reactive power injections, transformer tap positions, etc.). The PFSCs are often computed by knowing the compound admittance matrix of a given network and the grid states. However, when the branch parameters (or admittance matrix) are inaccurate or known with limited accuracy, the computed PFSCs can not be relied upon.
Uncertain PFSCs, when used in control, can lead to infeasible control set-points. In this context, this paper presents a method to quantify the uncertainty of the PFSCs from uncertain branch parameters and noisy grid-state measurements that can be used for formulating safe control schemes. We derive an analytical expression using the error-propagation principle. The developed tool is numerically validated using Monte Carlo simulations. 
\end{abstract}
\section{Introduction}
The non-linearities of the optimal power flow (OPF) are often tackled by its linearization, e.g., in \cite{jabr2019high} \textcolor{black}{where the derivatives of the controlled variables with respect to the control variables (referred to as power-flow sensitivity coefficients -- PFSCs) are used for expressing linear models of the grid constraints.} The PFSCs are used in key applications such as voltage control \cite{christakou2017voltage}, congestion management \cite{christakou2014gecn}, loss minimization, dispatching \cite{gupta2020grid}, etc. 
However, the computed PFSC might be uncertain due to inaccuracy in the grid parameters or \textcolor{black}{noisy} grid measurements. Therefore, to obtain a reliable control of \textcolor{black}{active distribution networks (ADNs)}, there is a need to quantify the \textcolor{black}{uncertainty on the PFSCs} such that those can be considered in the \textcolor{black}{uncertainty-aware} control formulation. 

In the existing literature, PFSC computation techniques can be broadly categorized into two types. The first is referred to as \emph{model-based schemes}, i.e., when the admittance parameters of the network are accurately known, for example, in \cite{zhou2008simplified, christakou2013efficient, maharjan2020enhanced}. The second referred to \emph{measurement-based control} schemes, use measurements to estimate the PFSCs, for example in \cite{mugnier2016model, da2019data, gupta2022model} etc. Both schemes may suffer from inaccuracies. When fed with uncertain admittance parameters, the model-based schemes lead to inaccurate and uncertain PFSCs. 
Similarly, the measurement-based schemes are prone to noises of the measuring units resulting in inaccurate estimates.  
Most existing works have focused on developing better estimation techniques for the PFSCs. Still, to the best of the author's knowledge, none of the schemes have quantified the amount of uncertainties on the computed/estimated PFSCs. \textcolor{black}{Currently,} the PFSC uncertainty is characterized by \textcolor{black}{computationally expensive} Monte-Carlo simulations \cite{christakou2017voltage} and then accounted for in the control problem. In \cite{gupta2022model}, the uncertainty is derived from a measurement-based \textcolor{black}{tool-chain} using nodal voltage and power measurements.

In this context, this work\footnote{\textcolor{black}{The paper is reusing the content from Author's doctoral dissertation \cite{GuptaThesis}.}} proposes a method to quantify uncertainty in PFSC computation. We refer to the case when PFSC is computed using the compound admittance matrix as in \cite{christakou2013efficient}. 
In particular, we derive an analytical \textcolor{black}{uncertainty propagation} tool using the \emph{principle of error propagation\footnote{\textcolor{black}{The term “error propagation” is widely used in the context of measurement instruments referring to their worst-case errors. This paper uses these error distributions of the measuring instruments along with errors on branch parameters to derive uncertainty on the computed PFSCs.}}} that is a function of admittance parameters, grid measurements and their respective uncertainties quantified by their variances and covariances. 
To the best of the author's knowledge, this is the first work quantifying uncertainty analytically from uncertain branch parameters in sensitivity coefficients. Compared to \cite{christakou2017voltage}, the proposed scheme is faster and straightforward.

The paper is organized as follows. Section~\ref{sec:prob_Stat} presents problem statement, Section~\ref{sec:method} recalls a method to compute PFSCs using compound admittance matrix, 
Section~\ref{sec:error_prop} presents the proposed uncertainty propagation method for PFSCs from uncertain compound admittance matrix, Section~\ref{sec:numberical_validation} presents the numerical validation of the proposed scheme, 
and finally, Section~\ref{sec:conclusions} summarizes the main findings of this work. 
\section{Problem Statement}\label{sec:prob_Stat}
We consider a generic power network for which we want to compute the PFSCs. For the sake of brevity, we limit the description to the computation of the voltage sensitivity coefficients and their uncertainties; other sensitivity coefficients can be similarly computed.
The objective is to assess the uncertainty of computed voltage sensitivity coefficients and their mean value by propagating the admittance parameter inaccuracies and grid measurements noise. The proposed tool is then numerically validated by the Monte Carlo approach for an IEEE benchmark network.
\section{Compound Admittance Matrix-based Sensitivity Coefficient Computation}
\label{sec:method}
In the following, we recall a method for computing the PFSCs using the compound admittance matrix, originally developed in \cite{christakou2013efficient} and extended in \cite{fahmy2021analytical}. 
The PFSC computation scheme relies on information on the compound admittance matrix and grid states. The method computes the PFSCs by solving a set of linear equations and guarantees a unique solution, given that the Jacobian of the load flow solution is invertible \cite{paolone2015static}. 

We assume a three-phase network configuration where the phases are denoted by $a, b$, and $c$, and let the number of phases be denoted by $p$, i.e., $p = 3$ in this case. Let the symbols $N_b$ and $\mathcal{N}_b = \{1,\dots, N_b\}$ refer to the number of buses/nodes and the set containing indices of the nodes in a generic network. 
Let the phase-to-ground nodal voltages be denoted\footnote{The complex quantity and its conjugate are denoted by $\moverbar{x}$ and $\underline{x}$, resp.} by $\moverbar{\boldsymbol{E}} = [ \moverbar{E}^{a}_1, \moverbar{E}^{b}_1, \moverbar{E}^{c}_1, \cdots, \moverbar{E}^{a}_{{N}_b}, \moverbar{E}^{b}_{{N}_b}, \moverbar{E}^{c}_{{N}_b}]$. 
Let the nodal apparent power injections are denoted by $\moverbar{\boldsymbol{S}} = [ \moverbar{S}^{a}_1, \moverbar{S}^{b}_1, \moverbar{S}^{c}_1, \cdots, \moverbar{S}^{a}_{{N}_b}, \moverbar{S}^{b}_{{N}_b}, \moverbar{S}^{c}_{{N}_b}]$, with $\moverbar{S}^{\phi}_i = P^{\phi}_i + jQ^{\phi}_i$ where 
\textcolor{black}{$P^{\phi}_i$ and $Q^{\phi}_i$ are active and reactive powers, respectively, for phase $\phi \in \{a,b,c\}$, node $i \in \mathcal{N}_b$; $\moverbar{S}^{\phi}_i $ is expressed as}
\begin{align}
\moverbar{S}^{\phi}_i =  \moverbar{E}^{\phi}_i \sum_{\substack{n \in \mathcal{N}_b\\ \phi' \in \{a,b,c\}}}\underline{Y}_{in}^{\phi\phi'} \underline{E}_{n}^{\phi'} &&  \textcolor{black}{\forall i\in \mathcal{N}_b, \phi \in \{a,b,c\}}
\label{eq:app_power}
\end{align} %
\normalsize
where, $\bar{Y}^{\phi, \phi'}_{in}$  ($\phi, \phi' \in \{a,b,c\}$) refers to an element of the admittance matrix referring to the line connecting between $i-$th and $n-$th nodes for in the compound admittance matrix $\bar{\mathbf{Y}}$ and mutual admittance between phases $\phi$ and $\phi'$.

Following the approach in \cite{christakou2013efficient, fahmy2021analytical}, we differentiate the equation in \eqref{eq:app_power} with respect to active power and reactive power  injections\footnote{Although the method proposed in \cite{fahmy2021analytical} is generic enough to apply to any kind of bus, we only consider voltage-independent and PQ buses in this work.} \textcolor{black}{$P^{\phi'}_l, Q^{\phi'}_l, l\in\mathcal{N}_b, \psi \in \{a,b,c\}$, it results in}

{\color{black}
\footnotesize
\begin{align}
    & \mathbb{1}_{ \substack{i = l, \\ \psi = \phi} } = \frac{{}\partial{\underline{E}^\phi_i}}{\partial{P_l^{\psi}}}\sum_{\substack{n \in \mathcal{N}_b\\ \phi' \in \{a,b,c\}}}{\bar{Y}}^{\phi\phi'}_{in}\moverbar{E}^{\phi'}_n + \underline{E}^{\phi}_i\sum_{\substack{n \in \mathcal{N}_b\\ \phi' \in \{a,b,c\}}}{\bar{Y}}^{\phi\phi'}_{in}\frac{{}\partial{\moverbar{E}^{\phi'}_n}}{\partial{P}^{\psi}_l}, \label{eq:der_coeffP}\\
    & -j\mathbb{1}_{ \substack{i = l, \\ \psi = \phi} } = \frac{{}\partial{\underline{E}^{\phi}_i}}{\partial{Q}^{\psi}_l}\sum_{\substack{n \in \mathcal{N}_b\\ \phi' \in \{a,b,c\}}}{\bar{Y}}^{\phi\phi'}_{in}\moverbar{E}^{\phi'}_n + \underline{E}^{\phi}_i\sum_{\substack{n \in \mathcal{N}_b\\ \phi' \in \{a,b,c\}}}{\bar{Y}}^{\phi\phi'}_{in}\frac{{}\partial{\moverbar{E}^{\phi'}_n}}{\partial{Q}^{\psi}_l} \label{eq:der_coeffQ}
\end{align}}
\normalsize
where $\frac{\partial{\square}}{\partial{\square}}$ refers to the partial derivatives. Here, $\mathbb{1}_{ \substack{i = l, \\ \psi = \phi} }$ denotes a vector of zeros with 1 when ${\{i = l\}}$ and ${\{\psi = \phi\}}$  are satisfied.
Rearranging the rows/columns of the eq.~\eqref{eq:der_coeffP}-\eqref{eq:der_coeffQ} for different nodes, the set of linear equations can be written as
\begin{align}
    \mathbf{z} = \mathbf{H}\mathbf{x} \label{eq:lin_sys}
\end{align}
where $\mathbf{H} \in \mathbb{R}^{2p(N_b-1)\times 2p(N_b-1)}$ and $\mathbf{z} \in \mathbb{R}^{{2p(N_b-1)} \times p(N_b-1)}$.
The elements of the matrix $\mathbf{H}$ and vector $\mathbf{z}$ are obtained using \cite{fahmy2021analytical}. 
Here, $\mathbf{x} \in  \mathbb{R}^{{2p(N_b-1)} \times p(N_b-1)}$ is the vector of unknown sensitivities, i.e., $\frac{{}\partial{\moverbar{E}^{\phi}_i}}{\partial{P}^{\phi'}_l}$ and $\frac{{}\partial{\underline{E}^{\phi}_i}}{\partial{P}^{\phi'}_l}$, where $i,l \in\mathcal{N}_b$ and $\phi, \phi' \in \{a,b,c \}$.
The $\mathbf{x}$ can be obtained by
\begin{align}
    \mathbf{x} = \mathbf{H}^{-1}\mathbf{z} \label{eq:lins_sense}.
\end{align}

It should be remarked that the expression in \eqref{eq:lins_sense} depends on the elements of the admittance matrix and the nodal voltage phasors. \textcolor{black}{As both the admittance matrix and nodal voltages are assumed to be uncertain, we need to propagate the error into the computation of the sensitivity coefficients. This is described next.}
\section{Uncertainty Propagation: From Uncertain Compound Admittance Matrix to the PFSCs}\label{sec:error_prop}
As anticipated, the elements of the $\mathbf{H}$ in eq.~\eqref{eq:lins_sense} might be uncertain due to the inaccuracy of the compound admittance matrix ($\bar{\mathbf{Y}}$) and the noise introduced by the measuring instruments. Therefore, the error on the $\bar{\mathbf{Y}}$ and grid voltage measurement $\bar{\boldsymbol{E}}$ needs to be propagated to evaluate the uncertainty on $\mathbf{x}$ in \eqref{eq:lins_sense}. The objective would be to compute the variance of the uncertainty coefficients contained in vector $\mathbf{x}$. 
\begin{hyp}
\label{hyp1}
    It is assumed that the measurement uncertainty on real and imaginary parts of $\bar{\boldsymbol{E}}$ can be modeled by Gaussian distribution, i.e., $\Re\{{\tilde{\bar{\boldsymbol{E}}}}\} \in \mathscr{N}(\Re\{{{\bar{\boldsymbol{E}}}}\}, \boldsymbol{\sigma}^{\Re}_{{{\bar{\boldsymbol{E}}}}})$ and $\Im\{{\tilde{\bar{\boldsymbol{E}}}}\} \in \mathscr{N}(\Im\{{{\bar{\boldsymbol{E}}}}\}, \boldsymbol{\sigma}^{\Im}_{{{\bar{\boldsymbol{E}}}}})$ with $\Re$ and $\Im$ denotes the real and imaginary parts of a complex number, $\tilde{\bar{\boldsymbol{E}}}$ the measurements of $\bar{\boldsymbol{E}}$ with noise, $ \boldsymbol{\sigma}^{\Re}_{{{\bar{\boldsymbol{E}}}}}$ and \textcolor{black}{$\boldsymbol{\sigma}^{\Im}_{{{\bar{\boldsymbol{E}}}}}$} the \textcolor{black}{standard (std.) deviations} on measurement uncertainty for $\Re\{{{\bar{\boldsymbol{E}}}}\}$ and $\Im\{{{\bar{\boldsymbol{E}}}}\}$, respectively. \textcolor{black}{Appendix~\ref{sec:projection} shows the expressions to compute real and imaginary variances for voltage phasors under a realistic noise model (from magnitudes and phase). Appendix~\ref{sec:noise_vald} validates the Gaussian assumption by quantile-quantile (QQ) plots.}
\end{hyp}
\begin{hyp}
\label{hyp2}
    It is assumed that the uncertainty on the real and imaginary parts of the elements of the compound admittance matrix can be modeled by Gaussian distributions.
\end{hyp}

\noindent \textbf{Remark:} In \eqref{eq:lins_sense}, the elements of $\mathbf{H}$ have multiplicative and additive elements consisting of elements of $\bar{\mathbf{Y}}$ and $\bar{\boldsymbol{E}}$ whose variances are known individually.
So, to compute the variance of elements of $\mathbf{H}$, we use propagation of uncertainty for \emph{multiplicative} and \emph{additive} error propagation rules. They are
\begin{subequations}
\begin{align}
    & \sigma(\Theta + \Phi)^2 \approx \sigma_{\Theta}^2 + \sigma_{\Phi}^2 + 2\sigma_{\Theta \Phi,}\\
    & \sigma(\Theta \times \Phi) \approx \Theta^2 \Phi^2(\sigma_{\Theta}^2/{\Theta}^2 + \sigma_{\Phi}^2/{\Phi}^2 + 2\sigma_{\Theta \Phi}/{\Theta \Phi}).
\end{align}
\end{subequations}

Let $[\sigma_{\mathbf{H}^{-1}}]$ be the \textcolor{black}{std. deviation} of matrix $\mathbf{H}^{-1}$, $[\sigma_{\mathbf{H}^{-1}}]_{ij}$ be the \textcolor{black}{std. deviation} of $ij-$th element of $\mathbf{H}^{-1}$, $[\sigma_{\mathbf{x}}]_{i}$ be the \textcolor{black}{std. deviation} of $i-$th element of $\mathbf{x}$, and $[\sigma_{\mathbf{z}}]_{j}$ be the variance of $j-$th element of $\mathbf{z}$.
Using the principle of error propagation, variance on $\mathbf{x}$ can be approximated as
\begin{subequations}
\label{eq:error_prop_set}
\begin{align}
    & [\sigma_{\mathbf{x}}]_{i}^2 = \sum_j\Big([\mathbf{H}^{-1}]^2_{ij}[\sigma_{\mathbf{z}}]_j^2 + {[\sigma_{\mathbf{H}^{-1}}]_{ij}^2}[\mathbf{z}]_j^2\Big). \label{eq:error_pr}
\end{align}

As $\mathbf{z}$ in \eqref{eq:lin_sys} is constant (consisting of 0's or 1's), $\sigma_{\mathbf{z}} = 0$, the expression in \eqref{eq:error_pr} reduces to
\begin{align}
    & [\sigma_{\mathbf{x}}]_{i}^2 = \sum_j{[\sigma_{\mathbf{H}^{-1}}]_{ij}^2}[\mathbf{z}]_j^2.\label{eq:error_pr2}
\end{align}

In \eqref{eq:error_pr2}, $[\sigma_{\mathbf{H}^{-1}}]$ is unknown, it can be computed by propagating the variance of the elements of $\mathbf{H}$ by following relation (as described in \cite{lefebvre2000propagation}), the self-correlation is given by
\begin{align}
    [\sigma_{\mathbf{H}^{-1}}]_{mn}^2 = \sum_i\sum_j [\mathbf{H}^{-1}]_{{mi}}^2[\sigma_{\mathbf{H}}]_{ij}^2[\mathbf{H}^{-1}]_{jn}^2 
\end{align}
and cross-correlation is given by
\begin{align}
\begin{aligned}
         & {\text{cov} (\mathbf{H}^{-1}_{mn}, \mathbf{H}^{-1}_{ab})}  = ~~~~~~~ \\
         & \sum_i\sum_j [\mathbf{H}^{-1}]_{mi}[\mathbf{H}^{-1}]_{ai}[\sigma_{\mathbf{H}}]_{ij}^2[\mathbf{H}^{-1}]_{jn}[\mathbf{H}^{-1}]_{jb}
\end{aligned}
\end{align}
\end{subequations}

\textcolor{black}{Since,} the matrix inversion is a non-linear operation, and the elements of the inverse matrix ($\mathbf{H}^{-1}$) are statistically correlated. Therefore, variance of each element of the matrix $\mathbf{H}^{-1}$ is of size $2p(N_b-1) \times 2p(N_b-1)$ itself by considering correlation of each element with every element in the matrix. Therefore, the size of $\sigma_{\mathbf{H}^{-1}}$ is $4p^2(N_b-1)^2 \times 4p^2(N_b-1)^2$.

\begin{table*}[t]
    \centering
        \caption{Performance comparison with the number of samples in Monte-Carlo simulations.}
    \begin{tabular}{|c|c|c|c|c|c|c|c|c|}
        \hline 
     &     & $\Re(\frac{\partial \bar{E}_3}{\partial P_2})$ & $\Re(\frac{\partial \bar{E}_3}{\partial P_4})$ & $\Re(\frac{\partial \bar{E}_4}{\partial P_4})$ & $\Im(\frac{\partial \bar{E}_3}{\partial P_2})$ & $\Im(\frac{\partial \bar{E}_3}{\partial P_3})$ & $\Im(\frac{\partial \bar{E}_4}{\partial P_4})$ & \textcolor{black}{Computation time (sec)}\\ 
         \hline
     Nominal value \textcolor{black}{(p.u)}         &     &  0.0071      &   0.0152      &   0.0239     &  0.0077      &   0.0176 &  0.0288 & \textcolor{black}{----}\\
             \hline
    \textcolor{black}{Std. deviation} (Analytical)  &    & 0.0006        &  0.0012       &   0.0018      & 0.0006        &    0.0011   & 0.0016& \textcolor{black}{0.15}\\
         \hline
             & $N_{\text{mc}}=$ 10 & 0.0009        &  0.0016       &   0.0021      &  0.0006      &   0.0012  &  0.0018 & \textcolor{black}{2.7}\\
 & $N_{\text{mc}}=$100      & 0.0007        &  0.0013      &   0.0020     &  0.0007      &   0.0013 &  0.0019 & \textcolor{black}{29.1}\\
   \textcolor{black}{Std. deviation} (Monte-Carlo)  & $N_{\text{mc}}=$1000      & 0.0007        &  0.0014       &   0.0020      &  0.0007      &   0.0013 &  0.0019 & \textcolor{black}{282.7}\\
       & $N_{\text{mc}}=$10000    & 0.0007        &  0.0014       &   0.0020      &  0.0007      &   0.0013 &  0.0019 & \textcolor{black}{2785}\\
            \hline
    \end{tabular}
    \label{tab:std_comparison_NMC}
\end{table*}

\begin{table*}[!htbp]
    \centering
        \caption{Performance comparison with respect to uncertainty on the admittance matrix \textcolor{black}{with $N_\text{mc}$ = 1000 and IT 1.0.}}
    \begin{tabular}{|c|c|c|c|c|c|c|c|}
        \hline
     &     & $\Re(\frac{\partial \bar{E}_3}{\partial P_2})$ & $\Re(\frac{\partial \bar{E}_3}{\partial P_4})$ & $\Re(\frac{\partial \bar{E}_4}{\partial P_4})$ & $\Im(\frac{\partial \bar{E}_3}{\partial P_2})$ & $\Im(\frac{\partial \bar{E}_3}{\partial P_3})$ & $\Im(\frac{\partial \bar{E}_4}{\partial P_4})$\\ 
         \hline
    $\sigma_{\bar{\mathbf{Y}}}$ (\% of $\bar{\mathbf{Y}}$) & Nominal value, \textcolor{black}{$x$ (p.u)}                  &  0.0071      &   0.0152      &   0.0239     &  0.0077      &   0.0176 &  0.0288\\
             \hline
   \multirow{2}{*}{0.5} & \textcolor{black}{$\sigma_{x,\text{MC}}$ (\% of $x$)}      & 0.337e-3 \textcolor{black}{(4.7\%)}        &  0.666e-3 \textcolor{black}{(4.4\%)}      &   0.986e-3 \textcolor{black}{(4.1\%)}     &  0.338e-3 \textcolor{black}{(4.4\%)}     &   0.659e-3 \textcolor{black}{(3.7\%)} &  0.938e-3 \textcolor{black}{(3.3\%)} \\
    & \textcolor{black}{$\sigma_{x,\text{AL}}$ (\% of $x$)}                             & 0.334e-3 \textcolor{black}{(4.7\%)}      &  0.662e-3  \textcolor{black}{(4.4\%)}    &   0.991e-3 \textcolor{black}{(4.1\%)}     & 0.316e-3 \textcolor{black}{(4.1\%)}      &    0.628e-3 \textcolor{black}{(3.6\%)}   & 0.900e-3 \textcolor{black}{(3.1\%)}\\
             \hline
    \multirow{2}{*}{1} & \textcolor{black}{$\sigma_{x,\text{MC}}$ (\% of $x$)}       & 0.0007 \textcolor{black}{(9.9\%)}      &  0.0014 \textcolor{black}{(9.2\%)}      &   0.0020 \textcolor{black}{(8.4\%)}     &  0.0007 \textcolor{black}{(9.1\%)}     &   0.0013 \textcolor{black}{(7.4\%)} &  0.0019 \textcolor{black}{(6.6\%)} \\
    & \textcolor{black}{$\sigma_{x,\text{AL}}$ (\% of $x$)}                         & 0.0006 \textcolor{black}{(8.5\%)}        &  0.0012 \textcolor{black}{(7.9\%)}      &   0.0018 \textcolor{black}{(7.5\%)}     & 0.0006 \textcolor{black}{(7.8\%)}       &    0.0011 \textcolor{black}{(6.3\%)}   & 0.0016 \textcolor{black}{(5.6\%)}\\
             \hline
    \multirow{2}{*}{2} & \textcolor{black}{$\sigma_{x,\text{MC}}$ (\% of $x$)}      & 0.0015  \textcolor{black}{(21.1\%)}      &  0.0030 \textcolor{black}{(19.7\%)}      &   0.0044 \textcolor{black}{(18.4\%)}     &  0.0014 \textcolor{black}{(18.2\%)}     &   0.0028 \textcolor{black}{(15.9\%)} &  0.0040 \textcolor{black}{(13.9\%)} \\
    & \textcolor{black}{$\sigma_{x,\text{AL}}$ (\% of $x$)}                        & 0.0013  \textcolor{black}{(18.3\%)}      &  0.0026 \textcolor{black}{(17.1\%)}      &   0.0033 \textcolor{black}{(13.8\%)}     & 0.0010 \textcolor{black}{(13\%)}       &    0.0021 \textcolor{black}{(11.9\%)}   & 0.0030 \textcolor{black}{(10.4\%)}\\
             \hline
    \end{tabular}
    \label{tab:std_comparison_NwYnoise}
\end{table*}

\section{Numerical Validation}
\label{sec:numberical_validation}
\textcolor{black}{We} numerically validate the proposed uncertainty estimation scheme using Monte Carlo simulations. The validation is performed \textcolor{black}{on} an IEEE benchmark network. 

\subsubsection{Simulation setup}
\begin{figure}[!h]
    \centering
    \includegraphics[width = 0.6\linewidth]{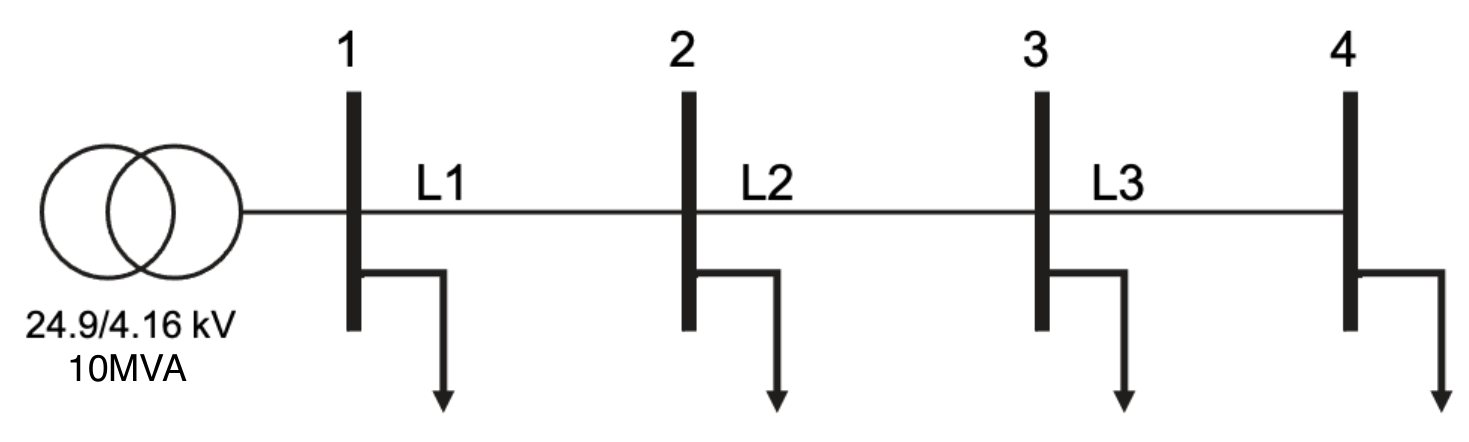}
    \caption{IEEE-4 balanced system.}
    \label{fig:4node}
\end{figure}

\textcolor{black}{The} proposed estimation scheme is validated on an IEEE-4 benchmark network \cite{kersting2001radial} (transposed line configuration) as shown in Fig.~\ref{fig:4node}.
The network is 24.9/4.16~kV, 10~MVA 3-phase operating in the balanced configuration.
The grid hosts 300~kW/150~kVar of active/reactive power demand at nodes 2, 3 and and 4 and photovoltaic generation of 480~kWp and 600~kWp at nodes 2 and 3.

For this performance evaluation, we use simulated \textcolor{black}{synchrophasor} noisy measurements \textcolor{black}{from phasor measurement units (PMUs)} on a network for which admittance parameters are known accurately. The noisy measurements are simulated as follows. First, load-flows are solved to compute the ground truth of the voltages phasors. Then, these phasors are added with noise in polar coordinates (i.e., in the magnitudes and phase). \textcolor{black}{For the noise, we assume independent and identically distributed (i.i.d) and Gaussian distribution. The measurement noises\footnote{\textcolor{black}{The noise introduced by IT is found to be more significant than PMU \cite{romano2014enhanced}, therefore the latter is neglected.}} are introduced in polar coordinates to reflect the real noise characteristics of instrument transformers (ITs). The ITs are characterized by magnitude and phase errors, specified in percentage for magnitudes and radians for phase angle. These values are characterized in \cite{IT_V} for different IT classes. The measurement noise is then projected on the rectangular coordinates to obtain $\boldsymbol{\sigma}^{\Re}_{{{\bar{\boldsymbol{E}}}}}, \boldsymbol{\sigma}^{\Im}_{{{\bar{\boldsymbol{E}}}}}$ as described in \cite{paolone2015static} as per {hypothesis}~\ref{hyp1}\textcolor{black}{, also described in Appendix~\ref{sec:projection}.}}

\begin{small}
\begin{algorithm}[!h]
{\color{black}
\caption{Monte-Carlo Simulations}\label{alg:MC_alg}
\begin{algorithmic}[1]
\Require {Nominal admittance: $\bar{\mathbf{Y}}$, 
nodal voltage phasor $\bar{\boldsymbol{E}}$}
\For{$k=1:N_\text{mc}$}
    \For{$n = 1:N_b$},
        \State $\delta_{\rho} = \mathscr{N}(0, {\sigma_{\rho}/3})$, $ \rho = |{\bar{E}_k(n)}| + \delta_{\rho}$, 
        \State $\delta_{\theta} = \mathscr{N}(0, {\sigma_{\theta}}/3)$, $\theta = \text{arg}({\bar{E}_k(n)}) + \delta_{\theta}$
        \State ${\tilde{\bar{E}}_k(n)} =  \rho e^{j\theta} \rightarrow \Re\{\tilde{\bar{E}}_k(n)\} + j\Im\{\tilde{\bar{E}}_k(n)\} $  
    \EndFor
    \For{$l = 1:N_b$},
        \For{$m = 1:N_b$}
        \State $\Delta Y^\text{re} = \mathscr{N}(0, \boldsymbol{\sigma}^{\Re}_{{{\bar{\boldsymbol{Y}}}}}(l,m))$
        \State $\Delta Y^\text{im} = \mathscr{N}(0, \boldsymbol{\sigma}^{\Im}_{{{\bar{\boldsymbol{Y}}}}}(l,m))$
        \State $\Re\{\tilde{\bar{\mathbf{Y}}}_k(l,m) \}= \Re\{{\bar{\mathbf{Y}}}(l,m)\} + \Delta Y^\text{re}(l,m)$
        \State $\Im\{\tilde{\bar{\mathbf{Y}}}_k(l,m) \}= \Im\{{\bar{\mathbf{Y}}}(l,m)\} + \Delta Y^\text{im}(l,m)$
    \EndFor
    \EndFor
        \State Compute sensitivity coefficient vector $\mathbf{x}$ using ${\tilde{\bar{\boldsymbol{E}}}_k}$ and ${\tilde{\bar{\mathbf{Y}}}_k}$ as expressed in \eqref{eq:lins_sense} and store in $\mathcal{X}(:,k)=\mathbf{x}$.
\EndFor
\State Compute standard deviation of each element of $\mathbf{x}$.
\end{algorithmic}}
\end{algorithm}
\end{small}
\subsubsection{Validation using the Monte-Carlo Simulations}
\label{app:Validation_error_MC}
The \textcolor{black}{uncertainty propagation} scheme developed in Sec.~\ref{sec:error_prop} is numerically validated by Monte-Carlo (MC) simulations. We use the IEEE 4-bus system as shown in Fig.~\ref{fig:4node}. 
The steps for MC simulation are described \textcolor{black}{Algorithm~\ref{alg:MC_alg}. The process includes noise sampling on voltage magnitudes and phase angles using IT class specifications in \cite{IT_V}, random error is introduced on the elements of the admittance matrix with chosen variance as listed in Table~\ref{tab:std_comparison_NwYnoise}. Then, the sensitivity coefficients are computed for each MC iteration and stored. Finally, variance of each sensitivity coefficient is computed.}

The variances computed using the analytical approach of \eqref{eq:error_pr2} are compared with the ones calculated using the MC. The comparison is shown in Table~\ref{tab:std_comparison_NMC}. It is shown for different $N_{\text{mc}}$. Here, we consider IT class 0.5 for the measurement noise on voltage phasors and 1.0~\% error on the admittance matrix. For the sake of brevity, the comparisons are only shown for six different coefficients, as listed in Table~\ref{tab:std_comparison_NMC}. \textcolor{black}{The coefficients are shown in per unit (p.u).}
As observed from Table~\ref{tab:std_comparison_NMC}, the standard deviation computed by the analytical approach and MC simulation matches for the simulations with $N_{\text{mc}}$ higher than 100. Hence, it can be concluded that the proposed analytical approach for \textcolor{black}{uncertainty propagation} works with good accuracy.

Furthermore, we vary the error of the admittance matrix as 0.5~\%, 1~\%, and 2~\%, respectively, to evaluate whether the \textcolor{black}{uncertainty propagation} scheme can give reliable estimates of the uncertainty on the sensitivity coefficients. This analysis is shown in Table~\ref{tab:std_comparison_NwYnoise}. \textcolor{black}{The symbols $\sigma_{x, \text{MC}}$ and $\sigma_{x, \text{AL}}$ represent the std. deviations computed using the Monte Carlo and analytical method, respectively.} As observed, it is concluded that the \textcolor{black}{uncertainty propagation} works well for 0.5~\% and 1~\%. \textcolor{black}{However, it differs slightly for the case of 2~\% error; the difference can be explained due to assumptions made on uncertainty propagation in Sec.~\ref{sec:error_prop} might not hold well for high standard deviations.}
\textcolor{black}{Besides, it should be noted that the with just 2\% of the uncertainty in the parameters led to uncertainty in sensitivity coefficients upto 22~\%; this is significant as the operating range for nodal voltages in a typical power system is limited to 10\% of the nominal operating point.}

\section{Conclusions}
\label{sec:conclusions}
In this work, we developed a tool to quantify the uncertainty of the power-flow sensitivity coefficients from uncertain compound admittance matrix and noisy grid measurements. The proposed scheme used the error propagation principle. The scheme was validated using Monte Carlo simulations on an IEEE benchmark network. 

The Monte Carlo validation confirmed that the proposed tool produces the same results provided that the uncertainty on the admittance parameters is not very high. Future work would be to improve uncertainty computation accuracy for a generic error distribution.
\appendix
{\color{black}
\subsubsection{Polar to Cartesian Noise Projection}
\label{sec:projection}
The standard deviations for real and imaginary parts (used for analytical uncertainty propagation) of the nodal voltages can be obtained by polar to Cartesian projection as described in \cite{paolone2015static}. It is
\begin{small}
\begin{align}
    & \begin{aligned} (\sigma^{\Re}_{\bar{E}})^2 = \frac{1}{2}\Big(1+\frac{\sigma_{\rho}^2}{9}\Big) & \rho^2(1+e^{-2\sigma^2_{\theta}}\cos(2\theta))\\ 
    + & \rho^2\cos^2(\theta)(1-2e^{-\frac{1}{2}{\sigma_{\theta}^2}})
    \end{aligned}
    \\
    & \begin{aligned} (\sigma^{\Im}_{\bar{E}})^2 = \frac{1}{2}\Big(1+\frac{\sigma_{\rho}^2}{9}\Big) & \rho^2(1+e^{-2\sigma^2_{\theta}}\cos(2\theta))\\
    + & \rho^2\sin^2(\theta)(1-2e^{-\frac{1}{2}{\sigma_{\theta}^2}})
    \end{aligned}
\end{align} 
\end{small}
where $\bar{E} = \rho e^{j\theta}$, $\sigma_{\rho}$ and $\sigma_{\theta}$ are the standard deviation of magnitude and phase in its polar representation.
}
{\color{black}
\subsubsection{Measurement noise validation}
\label{sec:noise_vald}
It presents verification of the hypothesis~\ref{hyp1}, i.e. whether the real and imaginary parts of the nodal voltages magnitudes can be approximated as Gaussian distributions. The QQ plot shown for IT 1.0 noise in Fig.~\ref{fig:QQplot} clearly overlaps with the normal line which concludes that the distributions can be approximated as Gaussian.
\begin{figure}[!htbp]
\centering
\subfloat[]{\includegraphics[width=0.43\columnwidth]{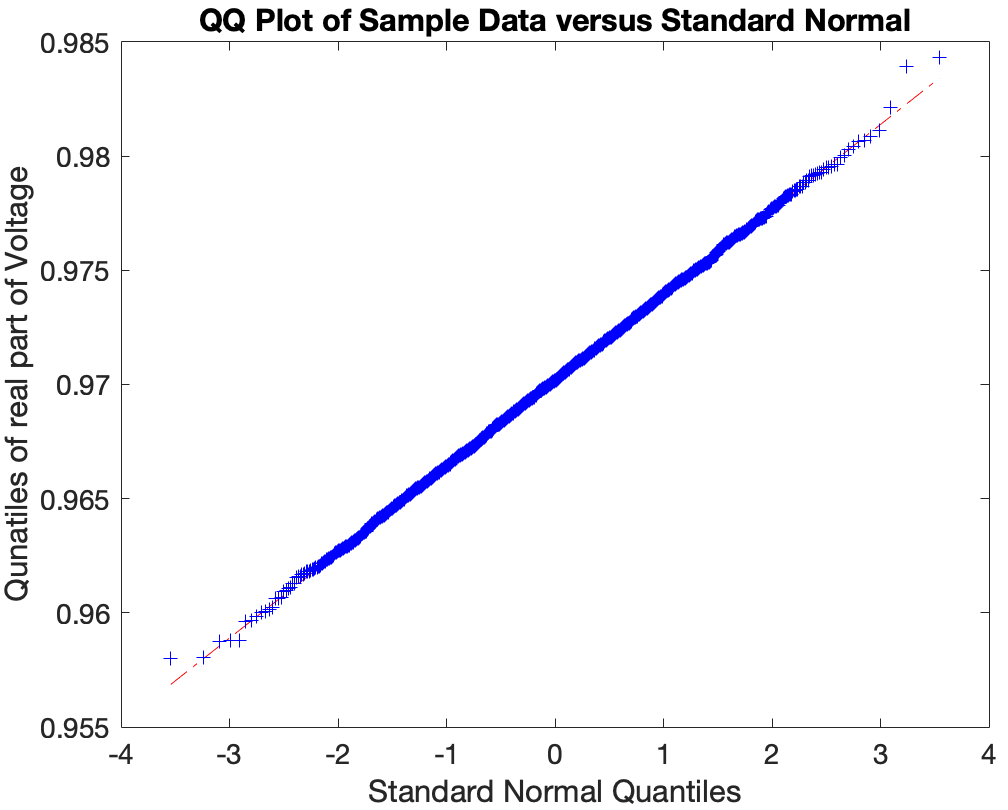}
\label{fig:qq_delP}}
\subfloat[]{\includegraphics[width=0.43\columnwidth]{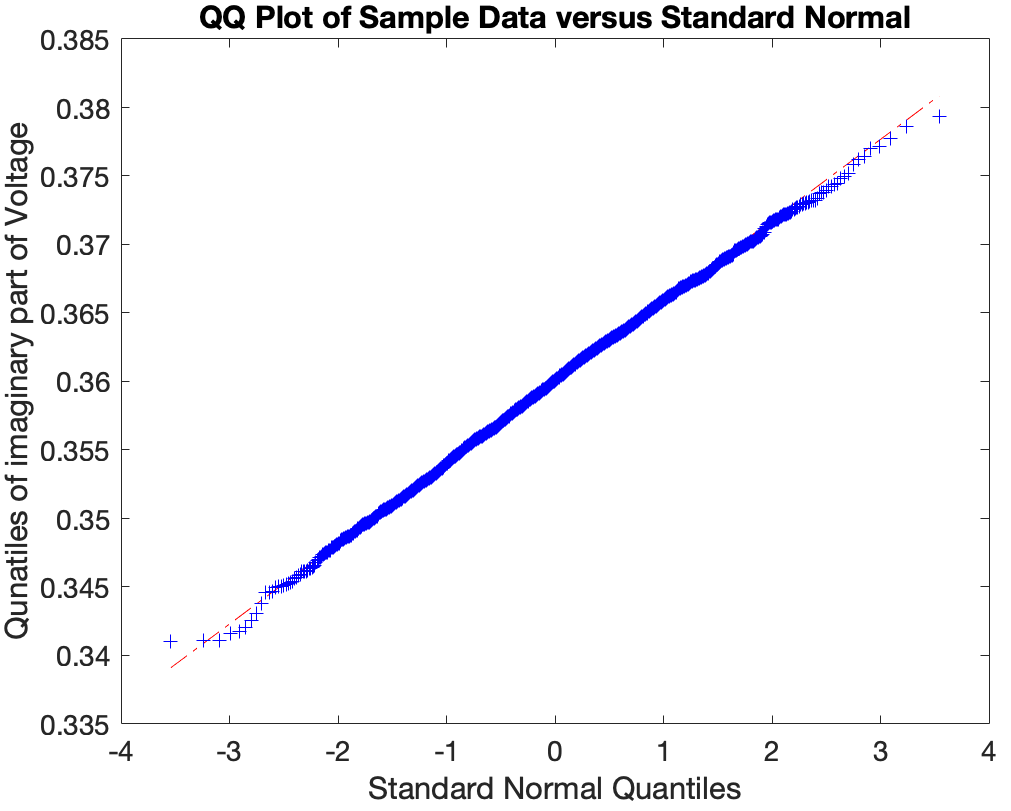}
\label{fig:qq_delQ}}
\caption{\textcolor{black}{QQ plot of the (a) real and (b) imaginary part of nodal voltage.}} \label{fig:QQplot}
\end{figure}
}

\section*{Acknowledgement}
\footnotesize
The author gratefully acknowledges the support of Prof. Mario Paolone, Head of the Distributed Electrical Systems Laboratory, EPFL, Switzerland for the useful feedback that he provided on the work presented in this paper.

\bibliographystyle{IEEEtran}
\bibliography{bibliography.bib}
\end{document}